\documentclass[journal]{IEEEtran}
\usepackage{cite}
\usepackage{amsmath,amssymb,amsfonts}
\usepackage{algorithmic}
\usepackage{graphicx}
\usepackage{textcomp}
\usepackage{adjustbox}
\usepackage{hyperref}
\usepackage{multirow}
\usepackage{multicol}

\DeclareMathOperator {\nb}{nb}

\begin{document}
\title{Ranking-Based Physics-Informed Line Failure Detection in Power Grids}
\author{Aleksandra Burashnikova, Wenting Li, Massih Amini,  Deepjoyti Deka, \IEEEmembership{Senior Member, IEEE}, Yury~Maximov, \IEEEmembership{Senior Member, IEEE},
\thanks{A. Burashnikova is with the University of Grenoble Alpes and Skolkovo Institute of Science and Technology, Moscow, 125043, Russia (e-mail: aleksandra.burashnikova@skoltech.ru). }
\thanks{M. Amini, is with the University Grenoble-Alpes, Grenoble, 38000 France.  (e-mail: massih-reza.amini@univ-grenoble-alpes.fr).}
\thanks{D. Deka, W. Li, and Y. Maximov are with 
the Theoretical Division, Los Alamos National Laboratory, Los Alamos, NM 
87545 USA (e-mails: $\{$deepjyoti,wenting,yury$\}$@lanl.gov ).}}

\maketitle

\begin{abstract}
Climate change increases the number of extreme weather events (wind and snowstorms, heavy rains, wildfires) that compromise power system reliability and lead to multiple equipment failures. Real-time and accurate detecting of potential line failures is the first step to mitigating the extreme weather impact and activating emergency controls. Power balance equations nonlinearity, increased uncertainty in generation during extreme events, and lack of grid observability compromise the efficiency of traditional data-driven failure detection methods. At the same time, modern problem-oblivious machine learning methods based on neural networks require a large amount of data to detect an accident, especially in a time-changing environment. This paper proposes a Physics-InformEd Line failure Detector (FIELD) that leverages grid topology information to reduce sample and time complexities and improve localization accuracy. Finally, we illustrate the superior empirical performance of our approach compared to state-of-the-art methods over various test cases. 
\end{abstract}


\section{Introduction}
\label{sec:introduction}
\IEEEPARstart{C}limate change and global warming result in increasing the number of extreme weather events \cite{sillmann2008indices} that compromise security and reliability of critical infrastructure (power and gas grids, telecommunications, transportation systems) \cite{birkmann2016extreme}.

According to the recent statistics of the National Center for Environmental Information\footnote{\url{https://www.ncdc.noaa.gov/billions/}}, the total cost of 310 recent major weather events exceeds 2.155 trillion dollars and is projected to increase in the near future~\cite{smith2013us}. Power grids are responsible for a substantial part of this~cost~\cite{stern2007economics,wiseman2016regional,silva2006fault}. 

One of the significant challenges in protecting a grid from impending a cascading blackout after a line failure is real-time localization of the faulted line followed by activating emergency controls \cite{begovic2005wide, zhang2016remedial}. Traditional data-driven methods for fault localization, such as traveling-wave \cite{parsi2020wavelet} and impedance-based ones \cite{aucoin1996high}, require high grid observability and sampling rates that are technically challenging and expensive for bulky systems~\cite{sundararajan2019survey}. Another line of algorithms leverages deep neural networks capabilities \cite{li2019real,li2021physicsA,zhang2020novel,misyris2020physics}; however, these methods suffer from high requirements on the amount of phasor-measurement unit data. The latter leads to the inability to make accurate and timely detection in a time-changing environment that is intrinsic for extreme weather events and, therefore, compromises power grid security.

\textbf{Contribution.} Our contribution is as follows. First, we propose Physics-InformEd Line failure  Detector (FIELD), a neural-network-based algorithm for detecting line faults in real-time. Particular advantages of our approach compared to conventional methods are its (1) higher accuracy and (2) lower data requirement achieved by leveraging power grid topology information. The latter reduces emergency control activation time and improves grid security posture. 

Second, for each line, we derive the conditional probability of being faulted. To this end, one can get a set of the most probable faulted lines and design ``umbrella'' control actions that guarantee the grid's stabilization if any of the above lines fail. Finally, we provide empirical support for the FIELD approach, demonstrating its superior performance over real and simulated data.
% \yury{Sasha, how many datasets do we have? Just one or a few? Real or simulated? Check the above pls.}
% \sasha{The data is simulated in the power system toolbox. See the statistics regarding simulated data in table \ref{tab:sim_grid_info}}

{\bf Paper structure.} The paper is organized as follows. Section~\ref{sec:setup} contains problem setup and provides necessary background information. Section \ref{sec:setup} describes our approach in detail and discusses its implications for emergency control in power systems. Section \ref{sec:algo} provides a critical review of existing results and indicates the role of our results in state-of-the-art. Empirical results and a short conclusion are given in Sections~\ref{ch5:exp} and \ref{ref:conc}, respectively. 

\section{Problem Setup and Background}\label{sec:setup}

\subsection{Notation.}
Let $E$, $|E| = m$, be a set of lines and $V$, $|V| = n$, is a set of buses in a power grid $G = \langle V, E\rangle$. Let $p,q \in\mathbb{R}^n$ be vectors of active and reactive power, $v\in\mathbb{R}^n$ be a vector of voltage magnitudes, and $\theta \in \mathbb{R}^n$ be a vector of voltage phases. We denote phase angle differences as $\theta_{ij}$, $(i,j)\in E$. Power grid buses consist of PQ (load) buses, PV (generation) buses, and a slack bus that often stands for the largest and slowest generator in the grid. We assume below that the phase angle~$\theta_i = 0$ for the slack bus $i$. 

The power grid is governed by the AC power flow equations and security constraints
\begin{subequations} 
    \begin{align}
        &p_k =\sum_{i=1}^{n_b}         v_iv_k\left(g_{ik}\cos{\theta_{ik}}+b_{ik}\sin{\theta_{ik}} \right)+g_\mathit{kk}v_k^2,\\
        & q_k =\sum_{l=1}^{n_b} v_iv_k\left(g_{ik}\sin{\theta_{ik}}-b_{ik}\cos{\theta_{ik}} \right)-b_\mathit{kk}v_k^2,\\
        & p^{\min} \le p_i \le p^{\max}, \; q^{\min} \le p_i \le q^{\max}\\
        & v_i^{\max} \le v_i \le v_i^{\min}, \theta_{ik}^{\min} \le \theta_{ik} \le \theta_{ik}^{\max},
    \end{align}
\end{subequations}
\noindent where network parameters $g_{ik}$ and $b_{ik}$ are susceptance and reactance of the transmission line connecting bus $i$ and $k$.

Paper notation is summarized in Table~\ref{tab:notation}.

\begin{table}[th]
    \centering
    \begin{adjustbox}{width=\linewidth}
    \begin{tabular}{l|l|l|l}
         $E$ & set of lines & $V$ & set of buses\\
         $m$ & number of lines& $n$ & number of buses\\
         $v$ & bus voltages, $v\in \mathbb{R}^n$ & $\theta$ & phase angles, $\theta\in\mathbb{R}^n$\\
         $p,q $ & \multicolumn{3}{l}{vector of active/reactive power injections} \\
         $d$ & number of PMUs & $V^d$ & set of nodes with PMUs\\
         $t$ & time index & & \\
         $y_{i}^t$ & \multicolumn{3}{l}{failure indicator at time $t$ at line $i$}\\
         $x^t$ & \multicolumn{3}{l}{a set of PMU measurements at time $t$, $(\{\theta^t_i, v^t_i\}_{i=1}^d)$} \\
         $\nb_E(\cdot), \nb_V(\cdot)$ & \multicolumn{3}{l}{list of adjacent edges, vertices}\\
         $\nb^k_E(\cdot), \nb^k_V(\cdot)$ & \multicolumn{3}{l}{$\nb^k_E(\cdot) = \underbrace{\nb_E((\dots \nb_E(\cdot))}_{k \text{ times}}, \nb^k_V(\cdot) = \underbrace{\nb_V(\dots \nb_V(\cdot))}_{k \text{ times}}$} \\
    \end{tabular}
    \end{adjustbox}
    \caption{Paper notation.}
    \label{tab:notation}
\end{table}

\subsection{Background.} Phasor Measurement Units (PMUs) enable high-resolution situational awareness of power grid state by providing information about voltage magnitude $v_i$, $i\in V$ and phase angle $\theta_{ij}$, $(i,j)\in E$ using a common time source for synchronization. Often PMUs are required at tap-changing transformers, complex loads, and PV (generation) buses. Despite the widespread PMUs and their role in grid monitoring, power grids remain covered only partly because of privacy and budget limitations. 

For notation simplicity, we assume w.l.o.g. that PMUs are placed at the first $d$ buses $V_d$ of the grid, $V_s\subseteq V$, and this placement does not change during the observation time. We refer $V_d$ as a set of observable buses. Furthermore, we receive a set of PMUs measurements $\boldsymbol{x^t} = (\{\theta_i^t, v_i^t\}_{i=1}^d)$ for each time~$t$, $0\le t \le T$. Let $\boldsymbol{y^t} \in \mathbb{R}^n$ be a an indicator of faulted lines, e.g. $y_{ij}^t = 1$ iff line $(i,j)$ is faulted at time $t$, $0\le t \le T$. 

The ability of PMU to measure the voltage phasor at the installed bus and the current phasor of all the branches connected to the PMU installed bus can help determine the remaining parameters to use for indirect measurements.

A particular advantage of PMU technology is the high sampling rate that dramatically increases situational awareness and detects grid failures in nearly real-time. For instance, for 60 Hz systems, PMUs must deliver between 10 and 30 synchronous reports per second, depending on the application. The timeline of the events in a power grid is described in Table~\ref{tab:my_label}.

\begin{table}[th]
    \centering
    \begin{tabular}{l|l}
        Event & time, sec. \\
        \hline\hline
         {\bf Transient Voltage Stability}& 0.2 -- 10 \\
         \hline 
         Line trip & 0.1 -- 1.5\\
         Static VAR Compensator (SVC) & 0.1 -- 1\\
         DC compensator & 0.1 -- 1\\
         Generator Inertial Dynamics & 0.5--5\\
         Undervoltage Load Shedding & 1--9\\
         Mechanically Switched Capacitors Dynamics & 0.15--2\\
         Generator/Excitation Dynamics & 0.15--3\\
         Induction Motor Dynamics & 0.1--2\\
         DC Converter LTCs & 4--20\\
         \hline
         {\bf Long-term Voltage Stability} & 20 -- 10000 \\
         \hline 
         Protective Relaying Including Overload Protection & 0.1 -- 1000\\
         Prime Mover Control & 1--100\\
         Auto-Reclosing & 15--150\\
         Excitation Limiting & 9--125\\
         Boiler Dynamics & 20--300\\
         Generator Change/AGC & 20--800\\
         Power Plant Operator & 40--1000\\
         Load Tap Changers and Dist, Voltage Reg. & 20--200\\
         System Operator & 60--10000\\
         RAS & 150--300\\
         RAP & 350--1000\\
         Gas Turbine Start-Up & 250--900\\
         Load Diversity/Thermostat & 200--2000\\
         Line/Transformer Overload & 600--2500\\
         Load/Power Transfer Increase & 250--7000 \\
         \hline
    \end{tabular}
    \caption{Events timeline in power grids.}
    \label{tab:my_label}
\end{table}

\section{Algorithm}\label{sec:algo}
To consider the topology of a power grid, we transform the binarized targets (fault or non-fault), that we used during training into two vectors: the first one includes the information about the faulted line and the second one consists of the information about the neighbors of the faulted line. In more details, suggest we have a sample $(\boldsymbol{\psi},\boldsymbol{y})$ with the features $\boldsymbol{\psi} \in R^d$, where $\boldsymbol{\psi}$ is some transformation over measurements $x^t$ and known parameters in power grid. Then the first vector of targets is defined as $\boldsymbol{y} = [y_1, \cdots,y_i, \cdots, y_n]^T \in R^n$, where in case of faulted line at the location $j$, $y_j =1$ and $y_i=0$ for $i!=j$. For the second vector of targets, let $nb_{E}(j)$ denote the neighborhood of the $j$th line, including the lines connected with $j$, and then $\hat{y}_i = 1/\text{nb}_{E}(j)$ only if $i \in \text{nb}_{E}(j)$. The definition of $\hat{y}$ is formalized at the equation defined at the Eq. \ref{eq:multilabel_targets}:

\begin{equation}
\label{eq:multilabel_targets}
\hat{y}(i) = 
\begin{cases}
1/|\text{nb}_{E}(j)|, & \text{if i} \in \text{nb}_{E}(j): \text{neighbor set of $j$} \\
0, & \text{else}\\
0 & i = j$~$ \text{the true location has weight 0}
\end{cases}
\end{equation}

\begin{figure}[th]
\centering
\vspace*{-10pt}
\includegraphics[width=0.85\columnwidth]{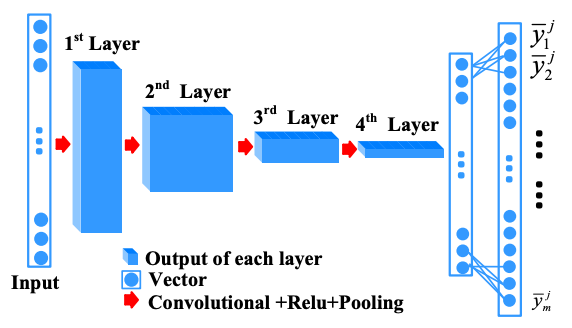}
\caption[Convolutional Network Architecture]{Architecture of the applied model proposed in \cite{DBLP:journals/corr/abs-1810-05247}.}
\label{fig:energy_arch}
\end{figure}

For the remains line target is equal to zero. Then the loss function $Loss(f(\boldsymbol{\psi}),\boldsymbol{y},\hat{\boldsymbol{y}})$ for the proposed model (architecture is presented on the Fig. \ref{fig:energy_arch}) over the samples $(\boldsymbol{\psi},\boldsymbol{y})$, where $f(\boldsymbol{\psi})$ are the predicting probabilities of the proposed model is defined as the sum of two terms of cross-entropy functions (here CE). The definition of CE is given below at the Eq. \ref{eq:cross entropy}:

\begin{equation}
CE(\boldsymbol{y},f(\boldsymbol{\psi})) = \sum_{i=1}^{n}y_i\cdot\log\left(\frac{\exp^{f_i(\boldsymbol{\psi})}}{\sum_{i=1}^{n}\exp^{f_{i}(\boldsymbol{\psi})}}\right)
\label{eq:cross entropy}
\end{equation}

Then, we could express the loss function more formally by the next Eq. \ref{eq:loss}:

\begin{equation}
Loss(\boldsymbol{y},\boldsymbol{\hat{y}},f(\boldsymbol{\psi})) = CE(\boldsymbol{y},f(\boldsymbol{\psi}))\cdot(1-\epsilon)+CE(\boldsymbol{\hat{y}},f(\boldsymbol{\psi}))\cdot \epsilon
\label{eq:loss}
\end{equation}

The architecture of the baseline model presented on the Fig. \ref{fig:energy_arch} is described in details in the paper of authors \cite{DBLP:journals/corr/abs-1810-05247}. It's suggested to use the convolution-based neural network with the information about the bus voltages and prepared features with a physical interpretation to make the predictions about fault location. To make the model more interpretable and to improve the output accuracy, we modified the loss function to the explained in the equation \ref{eq:loss} by including the network topology in the model and then provided an empirical evaluation of both approaches presented in section \ref{ch5:exp}.

\section{Empirical Study}\label{ch5:exp}

\subsection{Dataset}
To estimate the approaches we apply two benchmarks: SIM-LARGE and SIM-SMALL. SIM-SMALL was provided us by authors of \cite{DBLP:journals/corr/abs-1810-05247} for 68-bus power system.
The second dataset, SIM-LARGE, we simulated in the power system toolbox, based on nonlinear models \cite{PST}, a three-phase
short circuit fault lasting 0.2 seconds at line 5-6 in the IEEE 68-bus power system as in the SIM-SMALL. The main differences between the two benchmarks are the number of samples simulated for train, test, and validation sets, where the new simulated set is about ten times bigger. The second point is that the test set for SIM-LARGE is generated simultaneously for all fault types, as the train set for both datasets, whereas in the SIM-SMALL benchmark, there are separate test sets for each fault. This new simulation allows us to estimate the generalization property of the model to distinguish between different fault types. Also, it let us avoid the overfitting of the model on one particular class. 

The feature vector $\boldsymbol{\psi}$ is computed based on the idea lies in the baseline approach \cite{DBLP:journals/corr/abs-1810-05247}. Represented by the feature vectors, faulted lines in the power grid are then labeled by their locations. In the case of $m$ lines in the power grid, the number of output classes are equal to $m+1$, where an additional class is for the normal condition, which means there are no faults in the system. Below, the statistics regarding the size of simulated data for train, test, and validation evaluations are represented in the table \ref{tab:sim_grid_info}:

\begin{table}[ht!]
    \centering
    %\footnotesize
    \begin{tabular}{ccc}%ccccc
    \hline
    Dataset & Set & Size\\
    \hline
    \multirow{6}{*}{SIM-SMALL} & Train & 1210\\
    & \begin{tabular}{c}
    \hline
  TP - Test \\ 
  DLG - Test \\ 
  LG - Test \\ 
  LL - Test \\ \hline
\end{tabular} & \begin{tabular}{c}
    \hline
  71 \\ 
  71 \\ 
  70 \\
  71 \\ \hline
\end{tabular}\\
    & Validation & 1210\\
    \hline
   \multirow{3}{*}{SIM-LARGE} & Train & 14413 \\
    & Test & 994\\
    & Validation & 1207\\
    \hline
    \end{tabular}
    \caption{size of the train, test and validation parts.}
    \label{tab:sim_grid_info}
\end{table}

The fault cases provided in the data are simulated by changing the line impedance, depending on the type. For simulation, we consider a power grid of $n$ buses with a single line fault that may either be one of the following: three-phase short circuit (TP), a line to ground (LG), double line to ground (DLG), and line to line (LL) faults for SIM-SMALL and LG, DLG and LL for SIM-LARGE. To characterize the location of the faults in a power grid, the authors of \cite{DBLP:journals/corr/abs-1810-05247} propose to apply the substitution theory \cite{J.Quanyuan_2014} for deriving the equations related to pre- and during-fault system variables to express feature vectors. The feature vector $\boldsymbol{\psi} \in C^{n\times 1}$ based on the substitution theory is defined than in terms of the bus voltages variations $\Delta U$ before and during the faults and the admittance matrix $Y_0$ before the faults:

\begin{equation}
   {\psi}={\Delta U}\cdot Y_{0}
\end{equation}

Admittance matrix is an $n \times n$ matrix describing a linear power system with $n$ buses.
It represents the nodal admittance of the buses in a power system, where admittance measures how easily a circuit or device will allow a current to flow. The general mathematical expression of each element of the admittance matrix $Y_{ij}$ is represented as follows:

\begin{equation*}
Y_{ij} = 
 \begin{cases}
   y_i + \sum \limits_{k=1,...,n; k\neq i} y_{ki} &\text{i=j}\\
   -y_{ij} & i\neq j,
 \end{cases}
\end{equation*}
where $y_{ik}$ is the admittance between the bus $i$ and another bus $k$ connected to $i$. The term $y_i$ accounts for the admittance of linear loads connected to bus $i$ and the admittance-to-ground at bus $i$.
To understand the distribution of generated data, we provide statistics regarding the size of groups regarding the number of neighbors over lines. The results of the calculated statistics are introduced in Fig. \ref{fig:data_distr}:

\begin{figure}[th]
     \centering     \includegraphics[width=.45\columnwidth]{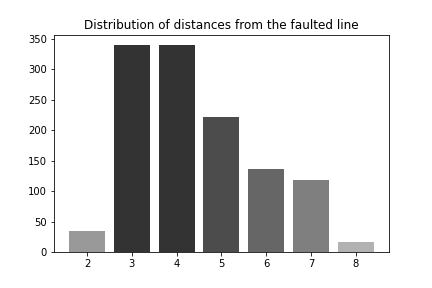}
    \includegraphics[width=.45\columnwidth]{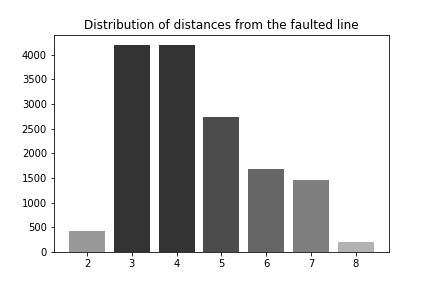}
     \caption{distribution of distances from a faulted line over the small (left) and the large(right) datasets.}
\label{fig:data_distr}
\end{figure}

\subsection{Signal to Noise Ratio}

SNR (signal-to-noise ratio) is a measure used in science and engineering that compares the level of the desired signal to the level of background noise. SNR is defined as the ratio between the output power of the transmitted signal and the power of the noise that distorts it.

\begin{equation}
    SNR = \frac{P_{signal}}{P_{noise}}=\frac{A^{2}_{signal}}{A^{2}_{noise}}
\end{equation}

$P$ here means average power, and $A$ is mean-square amplitude. Because many signals have a wide dynamic range, signals are often expressed using the logarithmic decibel scale. Then SNR ratio is expressed in decibels(dB) is transformed into the form:

\begin{equation}
    SNR_{dB} = 10\log_{10}\frac{P_{signal}}{P_{noise}}= 20\log_{10}\frac{A_{signal}}{A_{noise}}
\end{equation}

% \begin{equation}
%     SNR_{dB} = 20\log_{10}\frac{\sigma_{signal}}{\sigma_{noise}}
% \end{equation}

The ratio of SNR can take zero, positive or negative values. An SNR over 0 dB indicates that the signal level is greater than the noise level. The higher the ratio, the better the signal quality. The SNR of PMU measurements in different
regions can vary. We additionally explore this parameter over the test evaluations in subsection \ref{subsect:exp} of present chapter.

\subsection{Empirical Evaluation}
\label{subsect:exp}
The proposed model was trained using RMSProp optimizer, and for early-stopping criteria was suggested the next one: validation loss is computed over all validation data, then if $\min$ over the last 100 validation losses $<$ best loss, where the best loss is the minimum between the current best loss and the average over the last validation losses for 100 steps, then we continue to train, otherwise - stop. All the parameters such as learning rate, batch size, and the ratio that is responsible for how much information about the neighbors we take during the training and the remains parameters are set using cross-validation. To estimate the model, we apply an accuracy measure defined as the relation between the number of correctly detected faulted lines and the total number of faults. 
The first experiments are done on the small SIM-SMALL dataset over the full and partial observability cases. The partial measures range between $15\%$ and $30\%$ of buses and are estimated over 4 test sets for each fault class. The analysis of the results for two models could be found in the table \ref{tab: small_data_partial_measures}.

\begin{table*}[h!]
    \centering
    \resizebox{1.0\textwidth}{!}{
    \begin{tabular}{ccc|cc|cc|cc}
    \\
    \hline
    \multicolumn{1}{c}{} & \multicolumn{2}{c}{TP fault} & \multicolumn{2}{c}{DLG fault} & \multicolumn{2}{c}{LG fault}& \multicolumn{2}{c}{LL fault}
    \\
    \hline
    \% buses & No-neighbors & With-neighbors & No-neighbors & With-neighbors& No-neighbors & With-neighbors& No-neighbors & With-neighbors  \\
    \hline
100 & 98.59 & \bf{100.0} & \bf{100.0} & \bf{100.0} & \bf{100.0} & \bf{100.0} & \bf{100.0} & \bf{100.0}\\
30 & 91.55 & \bf{97.18} & 95.77 & \bf{98.59} & \bf{97.14} & \bf{97.14} & 98.59 & \bf{100.0}\\
25 & 78.87 & \bf{92.96} & 92.96 & \bf{97.18} & 94.29 & \bf{97.14} & 95.77 & \bf{98.59}\\
20 & 91.55 & \bf{94.36} & 90.14 & \bf{97.18} & 84.29 & \bf{94.29} & \bf{95.77} & \bf{95.77}\\
15 & 73.24 & \bf{88.73} & 95.77 & \bf{97.18} & 88.57 & \bf{92.86} & 88.73 & \bf{90.14}\\
\end{tabular}}
\caption{Comparison of the approaches based on the partial observability, SIM-SMALL data}
\label{tab: small_data_partial_measures}
\end{table*}

Based on the experiments, we see that the increase in grid observability improves the fault locations predictions quality. Also, it should be noticed that information about the grid topology also improves the final results on $2\%$-$18\%$ in comparison to the case without considering the neighbors during training in the loss function.

The results of the estimation the generalization property to distinguish the faults over different types are done on the SIM-LARGE dataset and presented in the table \ref{tab:SIM_BIG_diff_samples} for the range of train samples between $10$ and $100$ percentages with step $10$. For most cases, we could see the profit for the model with neighbors' topology. These results support the property of the generalization of the fault classes.

\begin{table*}[htb!]
    \centering
    \resizebox{.7\textwidth}{!}{
    \begin{tabular}{ccccccccccc}
    \\
    \hline
    +/- neighbors & 100 \% & 90\% & 80 \% & 70 \% & 60 \% & 50 \% & 40 \% & 30 \% & 20 \% & 10 \% \\
    \hline
 no neighbors & 95.07 & 93.86 & 93.66 & 92.76 & \bf{95.07} & 88.33 & \bf{91.44} & 93.16 & \bf{89.03} &  85.11\\
 with neighbors & \bf{95.57} & \bf{95.27} & \bf{95.07} & \bf{95.47} & 94.67 & \bf{91.95} & 90.74 & \bf{94.16} & 88.63 & \bf{89.64}  \\
\end{tabular}}
\caption{estimation for different sizes of the training set on SIM-LARGE data}
\label{tab:SIM_BIG_diff_samples}
\end{table*}

We also provide the experiments for partial bus observations for SIM-LARGE data similarly to the SIM-SMALL one. The results are presented in the table \ref{tab:sim_big_partial_measures}. Table~\ref{tab: small_data_partial_measures} illustrates that the ratio of measured buses and accuracy is also preserved in this case. 
The latter implies that the increase in the number of observations improves prediction accuracy for faulted line locations, which could be explained by the larger amount of input information provided for the model. Finally, our experiments indicate that the physics-informed model is less sensitive to the lack of observability than the physics-oblivious one. On average, the performance drop for the physics-informed model is 9\% compared to 13\% for the physics-oblivious one. 

\begin{table}[h!]
    \centering
    \resizebox{0.6\columnwidth}{!}{
    \begin{tabular}{ccc}
    \\
    \hline
    \multicolumn{3}{c}{LG fault}
    \\
    \hline
    \% buses & No-neighbors & With-neighbors\\
    \hline
100 & 95.07 & \bf{95.57}\\
30 & 85.41 & \bf{87.32}\\
25 & 78.27 & \bf{82.09}\\
20 & 79.48 & \bf{80.28}\\
15 & 71.93 & \bf{76.25}\\
\end{tabular}}
\caption{Comparison of the approaches based on partial observability on SIM-LARGE dataset}
\label{tab:sim_big_partial_measures}
\end{table}

\begin{figure}[ht]
\centering
\vspace*{-10pt}
\includegraphics[width=0.75\columnwidth]{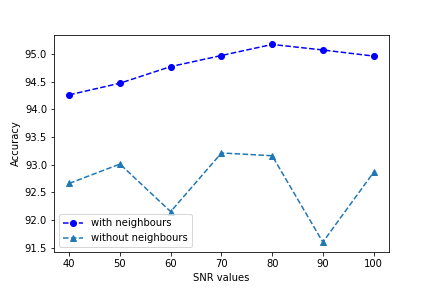}
\caption[SNR for various datasets]{Estimation of SNR approximation over the approaches on SIM-LARGE dataset. The results are provided for both models: with and without neighbors topology term in loss function}
\label{fig:snr_estimation}
\end{figure}

The test evaluations over the SNR parameter are done by ranging the approximation value of noise from 40dB to 100dB with the step size 10. The Gaussian noise of the same SNR was added both to the training and testing parts of the datasets. The structure of the CNN was kept the same but the hyperparameter as ratio $\epsilon$ in Eq. \ref{eq:loss} was additionally set up in
the noisy regime. Other parameters are the same. Results in Fig. \ref{fig:snr_estimation} indicate that the sensitivity
of both models to noise is different and that model based on neighbors topology is relatively more robust to the noise.

\subsection{U-Mann-Whitney Test}

\begin{figure}[ht]
\centering
\vspace*{-10pt}
\includegraphics[width=0.75\columnwidth]{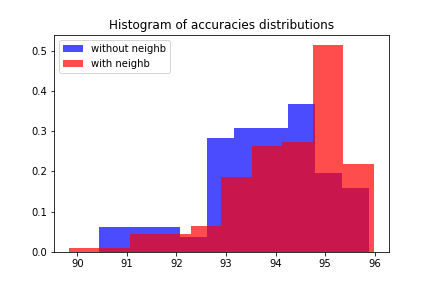}
\caption[Normalized histogram for samples distributions]{Normalized histogram for samples distributions. The histogram provides information about output accuracy for two models for different randomization.}
\label{fig:samples_dist_mannwhitney}
\end{figure}

Because of some instability in results from table \ref{tab:SIM_BIG_diff_samples}, we compare them based on the Mann-Whitney-Wilcoxon statistical criterion. The distributions of the output samples of accuracies are illustrated in Fig. \ref{fig:samples_dist_mannwhitney}. This U-criterion is used to assess the differences between two independent samples by the quantitative level of a feature. 

This method determines whether the zone of overlapping values between two rows is small enough. The lower the criterion's value, the more likely the differences between the parameter values in the samples are significant. U-Mann-Whitney test step-by-step:
\begin{itemize}
    \item To make a single ranked series from both compared samples, placing their elements according to the degree of increase and assigning a lower rank to a lower value with the number of elements in the first sample $n_1$ and $n_2$ in the second one.
    \item Divide a single ranked series into two, consisting of units of the first and second samples, respectively. Calculate the sum of ranks for each sample $R_1$ and $R_2$ separately, then calculate:
    \begin{equation}
        U_1 = n_1 \cdot n_2 + \frac{n_1 \cdot (n_1+1)}{2}-R_1
    \end{equation}
    
    \begin{equation}
        U_2 = n_1 \cdot n_2 + \frac{n_2 \cdot (n_2+1)}{2}-R_2
    \end{equation}
    \item Determine the value of the Mann-Whitney U-statistics by the formula $U=\max\{U_{1},U_{2}\}$.
    \item Using the table for the selected level of statistical significance, determine the critical value of the criterion for the data. Suppose the resulting value of $U$ is greater than or equal to the tabular one. In that case, it is recognized that there is a significant difference between the samples and an alternative hypothesis is accepted. The null hypothesis is accepted if the resulting value of $U$ is less than the table value.
\end{itemize}

In our case, as null hypothesis, we consider the equivalence of the mean for both samples, as an alternative hypothesis, we suggest that the mean of the model that takes into account the neighbours topology is greater than of the second one. The statistical significance, also denoted as $\alpha$, is the threshold probability of rejecting the null hypothesis when it is true. $p_{value}$ - is the actual probability (calculated from the resulting value of $U$) of rejecting the null hypothesis when it is true. So when $p_{value} < \alpha$, we assume that we reject the null hypothesis correctly.

The result of the Mann-Whitney statistical test is presented in %Fig. \ref{fig:mannwhitney_res}  for $60\%$ of training samples from SIM-BIG data (for all the remains ratios of the training data, the test was provided by analogy, and the results were the same). For the сomparison, a critical region of $2\sigma$ is given. 

\begin{figure}[ht]
\centering
\vspace*{-10pt}
\includegraphics[width=0.75\columnwidth]{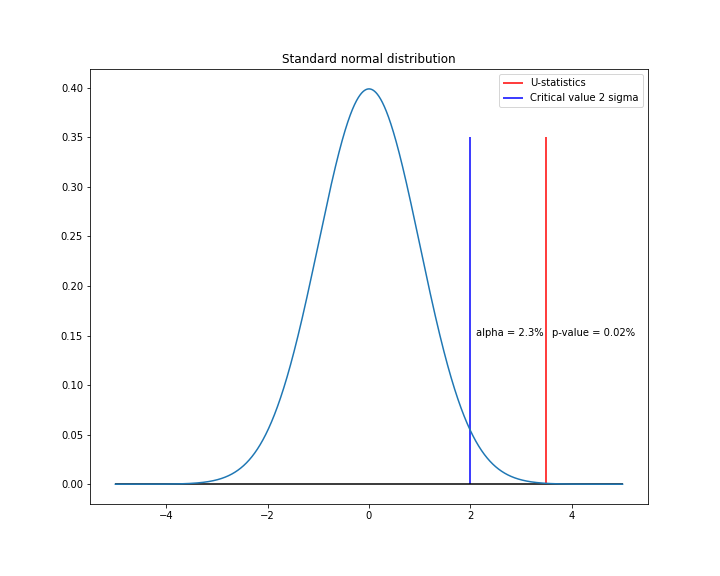}
\caption[The Mann-Whitney Statistics]{Computed Mann-Whitney Statistics for normal distribution.}
\label{fig:mannwhitney_res}
\end{figure}

Fig.~\ref{fig:mannwhitney_res} shows that the value of $p$-value is significantly less than alpha, so we reject the null hypothesis. Therefore, the mean of the model that considers the topology among neighbor lines in the power grid exceeds the mean of the baseline model. Thus we consider obtained results as statistically significant.

For power systems operational practice, it might be beneficial to present the solution in a simple logical form~\cite{boros2000implementation,hammer2006logical,maximov2013implementation} conventional for interpretation by a power system operator. 

\section{Conclusion}\label{ref:conc}
The problem of predicting power grid faults with a convolutional neural network is discussed in this chapter. Simulated datasets SIM-SMALL and SIM-BIG containing four and three types of errors were used to address the problem. We achieved the gains in accuracy by improving the loss function of the previously presented model \cite{DBLP:journals/corr/abs-1810-05247}. We added the term accounting neighbor information to the loss function to account for the neighbors of the line with a failure throughout the learning phase. To evaluate the statistical significance of the suggested technique, we used a statistical Mann-Whitney test to corroborate our findings. The test validated the approach's statical significance. Also, the modified model demonstrates its better robustness to noise conditions and partial observability. A similar approach can be used for the analysis of power generation reliability~\cite{stulov2020learning,mikhalev2020bayesian}.

\bibliographystyle{IEEEtran}
\bibliography{lcsys}

\end{document}